\documentclass[11pt]{article}
\pdfoutput=1
\usepackage{ACL2023}

\usepackage{times}
\usepackage{latexsym}
\usepackage[T2A,T1]{fontenc}
\usepackage[utf8]{inputenc}
\usepackage[russian,english]{babel}
\usepackage{tempora} % this supports Cyrillic
\usepackage{csquotes}
\usepackage{ragged2e}
\usepackage{tabularx}
\usepackage{graphicx}
\usepackage{multirow}
\usepackage{booktabs}
\usepackage{makecell}
\usepackage{microtype}

\usepackage{inconsolata}

\title{FERUZASPEECH: A 60 HOUR UZBEK READ SPEECH CORPUS WITH PUNCTUATION, CASING, AND CONTEXT}

\author{Anna Povey \\
  Redmond High School, WA, USA \\
  \texttt{annapovey@gmail.com} \\\And
  Katherine Povey \\
  University of Washington, USA \\
  \texttt{katherinepovey@gmail.com} \\}

\begin{document}

\maketitle

\begin{abstract}
    
This paper introduces FeruzaSpeech, a read speech corpus of the Uzbek language, containing transcripts in both Cyrillic and Latin alphabets, freely available for academic research purposes. This corpus includes 60 hours of high-quality recordings from a single native female speaker from Tashkent, Uzbekistan. These recordings consist of short excerpts from a book and BBC News. 
 This paper discusses the enhancement of the Word Error Rates (WERs) on CommonVoice 16.1's Uzbek data,  Uzbek Speech Corpus data, and FeruzaSpeech data upon integrating FeruzaSpeech.
\end{abstract}

\section{Introduction}

The Uzbek language, the official language of Uzbekistan, boasts upwards of 31 million native speakers across Central Asia. \footnote[1]{\url{https://www.worlddata.info/languages/uzbek.php}}Advancement in neural network models and deep learning have significantly improved automated speech recognition (ASR) and text-to-speech (TTS) technology in recent decades. Large freely available English datasets, such as LibriSpeech \cite{7178964}, Libriheavy \cite{kang2023libriheavy}, and GigaSpeech \cite{chen2021gigaspeech} are now more robust then ever, however, datasets for training these models in Uzbek are scarce.

In January 2023, the Uzbek government fully transitioned from using the Cyrillic alphabet to using the Latin alphabet
\footnote[2]{\url{https://interfax.az/view/826747}}, yet the country continues to use both alphabets. FeruzaSpeech is the first dataset to offer both Cyrillic and Latin transcription. To the best of our knowledge, FeruzaSpeech is also the only corpus to provide Cyrillic transcriptions at all. Datasets originally using Latin transcription cannot yet be accurately converted into Cyrillic text using online conversion calculators because there are a few discontinuities between the two alphabets. An example is when conversion calculators like this one\footnote[3]{\url{https://uzlatin.com/}} are used on Cyrillic text that include the soft sign \foreignlanguage{russian}{ь}, it is either lost or can be incorrectly reproduced becoming a hard sign \foreignlanguage{russian}{ъ}.

\begin{table}[th]
%\caption{Conversion Calculator on \foreignlanguage{russian}{Польша}}
  \caption{Conversion Calculator on \foreignlanguage{russian}{Польша}}
\begin{tabularx}{3.0in}{|X|X|}\hline
\textbf{Cyrillic to Latin} & \textbf{Latin to Cyrillic}\\\hline
\RaggedRight{\foreignlanguage{russian}{Пол{\bf ь}ша} -> Pol'sha}&
\RaggedRight{Pol'sha -> \foreignlanguage{russian}{Пол\textbf{ъ}ша}} \\\hline
\end{tabularx}
\end{table}

FeruzaSpeech aims to promote the development of speech recognition and speech synthesis technologies for the use of Uzbek speakers. Because this is a single speaker dataset with an absence of environmental noise it is better for STT when used in addition to other available speech corpuses. The dataset may be suitable for TTS applications, but such experiments are beyond the scope of this paper. It complements existing ASR datasets such Uzbek Speech Corpus (USC) \cite{musaev2021usc}, consisting of 105 hours from 958 speakers, and the Common Voice Uzbek Dataset \cite{ardila2019common} \footnote[4]{Download Page: \url{https://commonvoice.mozilla.org/en/datasets}}, with 265 hours from over 2,000 speakers. We chose these two corpuses because they were the only two other published datasets. When combined with these datasets, FeruzaSpeech enhances ASR model training. 

\section{FeruzaSpeech Corpus}
This section describes the layout of the FeruzaSpeech corpus metadata, transcription, and audio format. Instructions for downloading and utilizing the data can be found on HuggingFace. \footnote[5]{\url{https://huggingface.co/datasets/k2speech/FeruzaSpeech}}

\subsection{Dataset Type}
FeruzaSpeech consists of audio-book recordings from the texts of the book Choliqushi, a classic romance novel, and BBC Uzbek News read by our voice actress, Feruza. Table 2 shows the duration of each type of recording within the dataset. Initially read in the Cyrillic alphabet, the texts were converted to Latin using online tools\footnote[6]{ \url{https://www.lexilogos.com/keyboard/uzbek_conversion.htm} 
 and \url{https://uzlatin.com/}}, with some grammatical errors being manually fixed after the use of the conversion calculator. The final transcription provides Uzbek text in both the Cyrillic and Latin alphabets.
\begin{table}[h!]
  \caption{FeruzaSpeech Recordings}
  \label{tab:example}
  \centering
  \begin{tabular}{ r  r }
    \toprule
    \multicolumn{1}{c}{\textbf{Type}} & 
                                         \multicolumn{1}{c}{\textbf{Total}} \\
    \midrule
    Book &           $21.57$h~~~             \\
    BBC Uzbek &           $38.04$h~~~             \\
    Total &         $59.61$h~~~             \\
    \bottomrule
  \end{tabular}
\end{table}

\subsection{Evaluation and Training Sets}
FeruzaSpeech includes "Dev" (development), "Test" (testing), and "Train" (training) sets as detailed in Table 3. Both the Dev and Test sets only include BBC articles, while the Train set also includes the Choliqushi novel. 

\begin{table}[h!]
  \caption{FeruzaSpeech Sets}
  \label{tab:example}
  \centering
  \begin{tabular}{ r  r }
    \toprule
    \multicolumn{1}{c}{\textbf{Sets}} & 
                                         \multicolumn{1}{c}{\textbf{Total}} \\
    \midrule
    Dev &           $2.93$h~~~             \\
    Test &           $4.08$h~~~             \\
    Train &  $52.09$h~~~ \\
    \bottomrule
  \end{tabular}
\end{table}

\subsection{Audio Format}
The corpus contains high-quality, single-channel, 16-bit .wav audio files, available in 16kHz for ASR. The average recording length is 16.39 seconds, the minimum length is 3.78 seconds, and the maximum length is 50.69 seconds. Our segments are recordings of one to two full sentence and are much longer than the segments of USC \cite{musaev2021usc}, that are mostly 2 to 3 seconds. 

\begin{figure}[h!]
\centering
\includegraphics[scale=0.30]{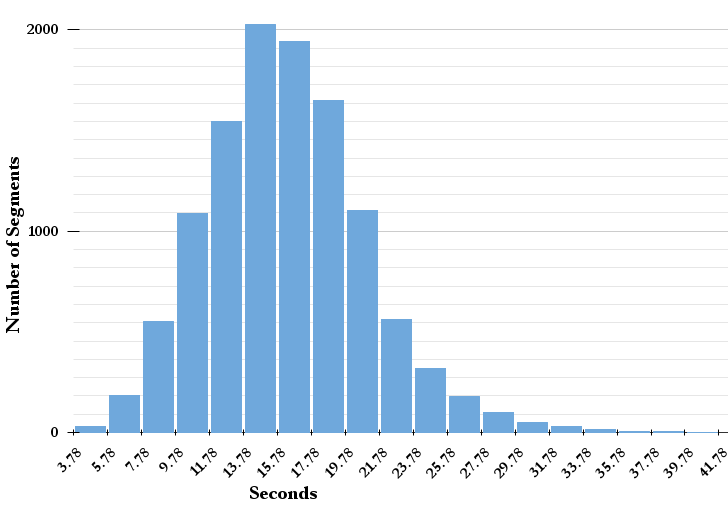}
\setlength{\belowcaptionskip}{13pt}
\caption{Length of FeruzaSpeech Segments}
\label{fig:x cubed graph}
\vspace*{-\baselineskip}
\end{figure}
\subsection{Sample Text}
Table 4 shows example excerpts from the CommonVoice and USC datasets in comparison to our proposed FeruzaSpeech dataset.

%\begin{center}
\begin{table}[h!]
  \caption{One Sentence of Sample Text from each of the Three Datasets with English Translation That Is Not in the Dataset for Reader}
\begin{tabularx}{3.1in}{|X|X|X|}\hline
\textbf{CommonVoice} & \textbf{USC} & \textbf{FeruzaSpeech}\\\hline
\RaggedRight{— Non dema! — dedi. — Nonni otini atama!}&
\RaggedRight{shundan so'ng u sen aytmasang men aytaman degandek qaradi}&
\RaggedRight{20 iyul kuni O‘zbekistonda 562 holatda kasallik qayd etilgan.}\\\hline
\RaggedRight{— Don't say bread! — he said. — Don't utter the word bread!}&
\RaggedRight{after that he looked like if you don't tell I will}&
\RaggedRight{On July 20, 562 cases of the disease were recorded in Uzbekistan.}\\\hline
\end{tabularx}
\vspace{-3mm}
\end{table}
%\end{center}
We can see that CommonVoice text was normalized but it has punctuation and casing, USC text is normalized to remove casing and punctuation, and FeruzaSpeech retains casing and punctuation. Regarding the choice to not normalize casing or punctuation, deep learning models have recently become powerful enough that for tasks like ASR and TTS it is now feasible to use "natural" text with no normalization. For instance, the recent E2TTS text-to-speech system \cite{eskimez2024e2ttsembarrassinglyeasy} is trained on data from Libriheavy \cite{kang2024libriheavy50000hoursasr} which is completely un-normalized. The use of un-normalized text for training tends to simplify speech processing systems because it could avoid the need for text normalization and inverse text normalization modules. Table 5 shows how the transcripts were provided in Latin and Cyrillic, but within this paper we only used Latin transcripts for comparison with available Latin datasets.

%\begin{center}

\begin{table}[h!]
  \caption{FeruzaSpeech Excerpt in Latin and Cyrillic with English Translation That Is Not in the Dataset for Reader}
\setlength\tabcolsep{3.0pt}
\begin{tabularx}{3.0in}{|X|X|}\hline
\textbf{FeruzaSpeech Latin} & \textbf{FeruzaSpeech Cyrillic} \\\hline
\RaggedRight{Ayni damda ishlatilib turilgan biometrik pasportlar 2019 yil 1 yanvardan deyarli yaroqsiz holatga keladi.}&
\RaggedRight{
\foreignlanguage{russian}{
Айни дамда ишлатилиб турилган биометрик паспортлар 2019 йил 1 январдан деярли яроқсиз ҳолатга келади.}
}\\\hline
\multicolumn{2}{|l|}{Biometric passports, which are currently in use,} \\
\multicolumn{2}{|l|}{will become almost useless from January 1,} \\
\multicolumn{2}{|l|}{2019.}\\ \hline
\end{tabularx}
\vspace{-3mm}
\end{table}

%{
%\texttt{}
%\begin{verbatim}

%common_voice_uz_28907218.mp3: Bugun ertalab Gyotenikiga taklifnoma oldim.

%common_voice_uz_28907221.mp3: Uning badiiy tasvir imkoniyatlarini rivojlantiradi

%1068-100.wav: 20 iyul kuni O‘zbekistonda 562 holatda kasallik qayd etilgan.

%1068-93.wav: Deylik, Buyuk Britaniyadagi har 10 kishidan to‘qqiztasida smartfon bor. Deyarli hamma smartfonga yopishib qolgan dunyoda endi undan "to‘ydim", deganlarning soni orta boshlabdi. Buni yaqinda o‘tkazilgan tadqiqot ko‘rsatdi. Qiziq...

%\end{verbatim}
%}

\section{Experiments}
To build our models we utilized the Next-Gen Kaldi framework and followed two recipes; the Icefall CommonVoice Stateless RNN-T Conformer model taken from the pruned\_transducer\_stateless7 recipe\footnote[7]{\url{https://github.com/k2-fsa/icefall/tree/master/egs/commonvoice/ASR/pruned_transducer_stateless7}} and the Librispeech  zipformer\footnote[8]{\url{https://github.com/k2-fsa/icefall/tree/master/egs/librispeech/ASR/zipformer}},  noting significant improvements in WER. These two models were selected because Icefall already contains CommonVoice scripts using
pruned\_transducer\_stateless7 in the French language and Librispeech zipformer is the current state of the art model in Next-Gen Kaldi. In our experiments we utilized three datasets: Common Voice 16.1 (CV), FeruzaSpeech (FS), and Uzbek Speech Corpus (USC).  All models were trained for 60 epochs. Table 6 outlines the duration of each training dataset.

\begin{table}[h!]
  \caption{Training Dataset Duration }
  \label{tab:example}
  \centering
  \begin{tabular}{ r  r }
    \toprule
    \multicolumn{1}{c}{\textbf{Datasets}} & 
                                         \multicolumn{1}{c}{\textbf{Training Duration}} \\
    \midrule
    CV &           $54.88$h~~~             \\
    FS &           $52.09$h~~~             \\
    USC &           $90.70$h~~~             \\
    CV + FS + USC &           $197.68$h~~~             \\
    \bottomrule
  \end{tabular}
\vspace{-4mm}
\end{table}

\subsection{Pruned-transducer-stateless7 Model}
The Stateless RNN-T Conformer model \cite{10094567} is a stateless transducer \cite{gulati2020conformer} with a conformer encoder that reduces memory consumption, and it outperformed the small zipformer model for every test set. All models in Tables 7 and 8 are trained with Casing and Punctuation. Table 7 presents the WERs when the model is scored with Casing and Punctuation, while Table 8 presents the WERs when the model is scored with Uppercase No Punctuation.

\begin{table}[!h]
\centering
\caption{The WERs of \textbf{Stateless RNN-T Conformer model} scored with \textbf{Casing and Punctuation (C\&P)} Common Voice 16.1 (CV), Uzbek Speech Corpus (USC), FeruzaSpeech (FS)}
\label{tab:comparsion-cases-punc}
%\adjustbox{maxwidth=\linewidth}{
\begin{tabular}{lllll}
\hline
\textbf{Method}          & \textbf{Dataset}       & \textbf{cv-test} & \textbf{fs-test}   & \textbf{usc-test}   \\ \hline
\multirow{2}{*}{\makecell{greedy \\ search}}   & CV           &  33.95    &  32.9   & 51.07 \\ 
                         & FS         &  89.54    &  11.58 &  85.67  \\
                         & CV+FS         &  32.49    &  9.93  & 46.89\\ 
                         & CV+FS+USC    &  29.91  &  9.79 &  12.05  \\\hline
\multirow{2}{*}{\makecell{modified \\ beam \\ search}}  & CV           &  31.98    &  31.88    & 51.61 \\  
                         & FS          &  89.10    &  11.25  & 85.22  \\ 
                         & CV+FS         &  30.47    &  9.85  & 48.6 \\
                         & CV+FS+USC         &  27.81    &  9.56  & 11.67\\ \hline

\end{tabular}
%\vspace{-4mm}
\end{table}

\begin{table}[h!]
\centering
\caption{The WERs of \textbf{Stateless RNN-T Conformer} model scored with \textbf{Uppercase No Punctuation (UNP)} Common Voice 16.1 (CV), Uzbek Speech Corpus (USC), and FeruzaSpeech (FS)}
\vspace{-2mm}
\label{tab:comparsion-cases-punc}
%\adjustbox{maxwidth=\linewidth}{
\begin{tabular}{lllll}
\hline
\textbf{Method}          & \textbf{Dataset}       & \textbf{cv-test} & \textbf{fs-test} & \textbf{usc-test} \\ \hline
\multirow{2}{*}{\makecell{greedy \\ search}}   & CV           &  21.03    &  20.15 & 35.11   \\ 
                         & FS         &  87.18    &  5.85  & 77.78\\
                         & CV+FS         &  18.91    & 4.44 & 30.53 \\
                         & CV+FS+USC         &  12.07    &  4.17 & 12.05\\ \hline
\multirow{2}{*}{\makecell{modified\\ beam \\ search}}  & CV           & 20.16    &  19.34  & 34.03    \\  
                         & FS          & 86.26    &  5.50  & 76.24  \\ 
                         & CV+FS         &  18.33    &  4.24  & 29.67\\
                         & CV+FS+USC         &  11.17    &  4.05 & 11.67\\ \hline

\end{tabular}
\vspace{-4mm}
\end{table}

\subsection{Zipformer Model}

We followed a similar procedure for the zipformer model as we did for the Stateless RNN-T Conformer model. This time, we trained a separate model on each of the following three datasets: CV, FS, CV+FS. This differs from the previous section because we excluded the USC training set. Once again, we recorded the WER for each model when
tested on each of the following test sets: cv-test, fs-test, and usc-
test, sharing results for both the greedy search and modified beam search as methods of decoding. All models in Table 9 and 10 are trained with Casing and Punctuation. Table 9 presents the WERs when the model is scored with Casing and Punctuation, while Table 10 presents the WERs when the model is scored with Uppercase No Punctuation.

Note that the Common Voice recipe with default settings in the Icefall project wasn't converging for zipformer \cite{kang2023libriheavy}, so we used "small zipformer" \footnote[9]{\url{https://github.com/k2-fsa/icefall/blob/master/egs/librispeech/ASR/RESULTS.md}} \cite{yao2024zipformer} parameters to account for the size of our datasets.

\begin{table}[!h]
\centering
\caption{The WERs of \textbf{zipformer} model scored with \textbf{Casing and Punctuation (C\&P)} Common Voice 16.1 (CV) and FeruzaSpeech (FS)}
\vspace{-2mm}
\label{tab:comparsion-cases-punc}
%\adjustbox{maxwidth=\linewidth}{
\begin{tabular}{lllll}
\hline
\textbf{Method}          & \textbf{Dataset}       & \textbf{cv-test} & \textbf{fs-test} & \textbf{usc-test} \\ \hline
\multirow{2}{*}{\makecell{greedy \\ search}}   & CV           &   37.00  &  34.54 &53.4  \\ 
                         & FS         &  93.09    &  14.32  & N/A  \\
                         & CV+FS         &  35.90    &  11.05 &52.86  \\ \hline
\multirow{2}{*}{\makecell{modified\\ beam \\ search}}  & CV        &  33.96    &  32.41   &54.07  \\  
                         & FS          &  92.61    &  13.28   &N/A   \\ 
                         & CV+FS         &  33.15    &  10.75 & 53.08 \\ \hline

\end{tabular}
\vspace{-4mm}
\end{table}

\begin{table}[!h]
\centering
\caption{The WERs of \textbf{zipformer} model scored with \textbf{Uppercase No Punctuation (UNP)} Common Voice 16.1 (CV) and FeruzaSpeech (FS)}
\vspace{-2mm}
\label{tab:comparsion-cases-punc}
%\adjustbox{maxwidth=\linewidth}{
\begin{tabular}{lllll}
\hline
\textbf{ Method}          & \textbf{Dataset}       & \textbf{cv-test} & \textbf{fs-test}  &\textbf{usc-test}\\ \hline
\multirow{2}{*}{\makecell{greedy \\ search}}   & CV           &  23.01    &  20.94  &38.75  \\ 
                         & FS         &  91.34    &  8.97 & N/A  \\
                         & CV+FS         &  22.34    & 5.45 &  37.4\\ \hline
\multirow{2}{*}{\makecell{modified \\ beam \\ search}}  & CV           & 21.92    &  20.21  & 37.44   \\  
                         & FS          & 90.54    &  7.88  & N/A   \\ 
                         & CV+FS         &  21.35    &  5.10 & 35.94 \\ \hline

\end{tabular}
\vspace{-4mm}
\end{table}
\section{Results}

When adding the FeruzaSpeech dataset to the CommonVoice16.1 dataset while training the Stateless RNN-T Conformer model, WER improved 1.49 to 2.12 percent absolutely on cv-test and 3.01 to 4.58 percent absolutely on usc-test in Tables 7 and 8. And for the Zipformer model, WER improved 0.57 to 1.1 percent absolutely on cv-test and 0.54 to 1.5 percent absolutely on usc-test in tables 9 and 10.  This shows that FeruzaSpeech contains quality data and is a useful addition to the current public library of Uzbek speech corpuses for TTS applications.
Also, the paper presenting the USC dataset \cite{musaev2021usc} reports that the usc-test had a WER of 17.4\%. Our best result for the usc-test WER is 11.67\%, which is an improvement of 5.73\%. 
According to Table 6 and 7, when a Stateless RNN-T Conformer model was built using all three datasets combined: CV, FS, and USC, and using modified beam search as the decoding method, the model produced the best WERs for every test. Our best recorded WER on the Common Voice test set is 11.17\%, as shown in Table 7. The best WER for the FeruzaSpeech test set is 4.05\%, and the best WER for the Uzbek Speech Corpus test is 11.67\%.

\section{Conclusion}
The development of FeruzaSpeech is a significant step forward in the field of Uzbek speech technology. By offering a dual alphabet corpus, this project bridges the gap between the use of Cyrillic and Latin scripts for Uzbek speakers. Our work also highlights the need for accurate alphabet conversion tools, specifically for more nuanced aspects of the language such as the soft sign (\foreignlanguage{russian}{ь}), which tends to be lost in translation from Cyrillic to Latin.

Through integrating FeruzaSpeech with existing Uzbek datasets, notable improvements in WERs were demonstrated. In the future, we will provide this same data in a higher sampling rate and bit depth that will be more suitable for TTS. Since we recognize the value of continuity in voice data for TTS applications, our future endeavors will also focus on expanding this corpus with additional recordings from the same native speaker. This strategy aims to enrich the dataset with consistent voice quality and style across the corpus which is essential for developing TTS models.

In sum, FeruzaSpeech is beneficial for ASR model enhancement when used in addition to existing Uzbek language datasets, as observed in WER improvements.  Applications of this dataset for TTS will also be explored.
\section{Limitations}
FeruzaSpeech is not an effective stand alone corpus for STT applications and should be used in compliment with other corpuses such as the Common Voice Uzbek Dataset and Uzbek Speech Corpus explored above. FeruzaSpeech has an average segment length of 16.39 seconds which each contain one or two full sentences which could be segmented into shorter utterances. The audio has no background noise and contains a singular female speaker which is not optimal for STT. 
\bibliography{anthology,custom}

\begin{thebibliography}{10}
\expandafter\ifx\csname natexlab\endcsname\relax\def\natexlab#1{#1}\fi

\bibitem[{Ardila et~al.(2019)Ardila, Branson, Davis, Henretty, Kohler, Meyer, Morais, Saunders, Tyers, and Weber}]{ardila2019common}
Rosana Ardila, Megan Branson, Kelly Davis, Michael Henretty, Michael Kohler, Josh Meyer, Reuben Morais, Lindsay Saunders, Francis~M Tyers, and Gregor Weber. 2019.
\newblock Common voice: A massively-multilingual speech corpus.
\newblock \emph{arXiv preprint arXiv:1912.06670}.

\bibitem[{Chen et~al.(2021)Chen, Chai, Wang, Du, Zhang, Weng, Su, Povey, Trmal, Zhang, Jin, Khudanpur, Watanabe, Zhao, Zou, Li, Yao, Wang, Wang, You, and Yan}]{chen2021gigaspeech}
Guoguo Chen, Shuzhou Chai, Guanbo Wang, Jiayu Du, Wei-Qiang Zhang, Chao Weng, Dan Su, Daniel Povey, Jan Trmal, Junbo Zhang, Mingjie Jin, Sanjeev Khudanpur, Shinji Watanabe, Shuaijiang Zhao, Wei Zou, Xiangang Li, Xuchen Yao, Yongqing Wang, Yujun Wang, Zhao You, and Zhiyong Yan. 2021.
\newblock \href {http://arxiv.org/abs/2106.06909} {Gigaspeech: An evolving, multi-domain asr corpus with 10,000 hours of transcribed audio}.

\bibitem[{Eskimez et~al.(2024)Eskimez, Wang, Thakker, Li, Tsai, Xiao, Yang, Zhu, Tang, Tan, Liu, Zhao, and Kanda}]{eskimez2024e2ttsembarrassinglyeasy}
Sefik~Emre Eskimez, Xiaofei Wang, Manthan Thakker, Canrun Li, Chung-Hsien Tsai, Zhen Xiao, Hemin Yang, Zirun Zhu, Min Tang, Xu~Tan, Yanqing Liu, Sheng Zhao, and Naoyuki Kanda. 2024.
\newblock \href {http://arxiv.org/abs/2406.18009} {E2 tts: Embarrassingly easy fully non-autoregressive zero-shot tts}.

\bibitem[{Gulati et~al.(2020)Gulati, Qin, Chiu, Parmar, Zhang, Yu, Han, Wang, Zhang, Wu, and Pang}]{gulati2020conformer}
Anmol Gulati, James Qin, Chung-Cheng Chiu, Niki Parmar, Yu~Zhang, Jiahui Yu, Wei Han, Shibo Wang, Zhengdong Zhang, Yonghui Wu, and Ruoming Pang. 2020.
\newblock \href {http://arxiv.org/abs/2005.08100} {Conformer: Convolution-augmented transformer for speech recognition}.

\bibitem[{Kang et~al.(2023{\natexlab{a}})Kang, Guo, Kuang, Lin, Luo, Yao, Yang, Żelasko, and Povey}]{10094567}
Wei Kang, Liyong Guo, Fangjun Kuang, Long Lin, Mingshuang Luo, Zengwei Yao, Xiaoyu Yang, Piotr Żelasko, and Daniel Povey. 2023{\natexlab{a}}.
\newblock \href {https://doi.org/10.1109/ICASSP49357.2023.10094567} {Fast and parallel decoding for transducer}.
\newblock In \emph{ICASSP 2023 - 2023 IEEE International Conference on Acoustics, Speech and Signal Processing (ICASSP)}, pages 1--5.

\bibitem[{Kang et~al.(2023{\natexlab{b}})Kang, Yang, Yao, Kuang, Yang, Guo, Lin, and Povey}]{kang2023libriheavy}
Wei Kang, Xiaoyu Yang, Zengwei Yao, Fangjun Kuang, Yifan Yang, Liyong Guo, Long Lin, and Daniel Povey. 2023{\natexlab{b}}.
\newblock Libriheavy: a 50,000 hours asr corpus with punctuation casing and context.
\newblock \emph{arXiv preprint arXiv:2309.08105}.

\bibitem[{Kang et~al.(2024)Kang, Yang, Yao, Kuang, Yang, Guo, Lin, and Povey}]{kang2024libriheavy50000hoursasr}
Wei Kang, Xiaoyu Yang, Zengwei Yao, Fangjun Kuang, Yifan Yang, Liyong Guo, Long Lin, and Daniel Povey. 2024.
\newblock \href {http://arxiv.org/abs/2309.08105} {Libriheavy: a 50,000 hours asr corpus with punctuation casing and context}.

\bibitem[{Musaev et~al.(2021)Musaev, Mussakhojayeva, Khujayorov, Khassanov, Ochilov, and Atakan~Varol}]{musaev2021usc}
Muhammadjon Musaev, Saida Mussakhojayeva, Ilyos Khujayorov, Yerbolat Khassanov, Mannon Ochilov, and Huseyin Atakan~Varol. 2021.
\newblock Usc: An open-source uzbek speech corpus and initial speech recognition experiments.
\newblock In \emph{Speech and Computer: 23rd International Conference, SPECOM 2021, St. Petersburg, Russia, September 27--30, 2021, Proceedings 23}, pages 437--447. Springer.

\bibitem[{Panayotov et~al.(2015)Panayotov, Chen, Povey, and Khudanpur}]{7178964}
Vassil Panayotov, Guoguo Chen, Daniel Povey, and Sanjeev Khudanpur. 2015.
\newblock \href {https://doi.org/10.1109/ICASSP.2015.7178964} {Librispeech: An asr corpus based on public domain audio books}.
\newblock In \emph{2015 IEEE International Conference on Acoustics, Speech and Signal Processing (ICASSP)}, pages 5206--5210.

\bibitem[{Yao et~al.(2024)Yao, Guo, Yang, Kang, Kuang, Yang, Jin, Lin, and Povey}]{yao2024zipformer}
Zengwei Yao, Liyong Guo, Xiaoyu Yang, Wei Kang, Fangjun Kuang, Yifan Yang, Zengrui Jin, Long Lin, and Daniel Povey. 2024.
\newblock \href {http://arxiv.org/abs/2310.11230} {Zipformer: A faster and better encoder for automatic speech recognition}.

\end{thebibliography}
\bibliographystyle{acl_natbib}

\end{document}